\pdfoutput=1
\documentclass[
 pra, reprint,
 superscriptaddress,
 amsmath,amssymb,
 aps,longbibliography
]{revtex4-2}

\usepackage{graphicx}
\usepackage{amsmath}
\usepackage{array}
\usepackage{xcolor}
\usepackage[bookmarks=false]{hyperref}
\usepackage[normalem]{ulem}
\usepackage{soul}
\usepackage[range-phrase = --]{siunitx}
\usepackage{fancyhdr}
\usepackage{duckuments}
\usepackage{nicematrix}
\NiceMatrixOptions{cell-space-limits = 2pt}
\usepackage[caption=false]{subfig}
\newcommand{\slh}[3]{\left( \mathopen{}#1\mathclose{},\, \mathopen{}#2\mathclose{},\, \mathopen{}#3\mathclose{} \right)}

\renewcommand{\Re}{\operatorname{Re}}
\renewcommand{\Im}{\operatorname{Im}}

\newcommand{\ket}[1]{\mathinner{|#1\rangle}}

\newcommand{\mi}{\ensuremath{{i}}} 

\begin{document}

\title{Transmon Architecture for Emission and Detection of Single Microwave Photons}

\author{Daniel L. Campbell}
\email{Daniel.Campbell.22@us.af.mil}
\affiliation{Air Force Research Laboratory, Information Directorate, Rome NY 13441 USA}

\author{Stephen McCoy}
\affiliation{Booz Allen Hamilton, McLean, VA USA 22012}

\author{Melinda Andrews}
\affiliation{Booz Allen Hamilton, McLean, VA USA 22012}

\author{Alexander Madden}
\affiliation{Booz Allen Hamilton, McLean, VA USA 22012}

\author{Viva R. Horowitz}
\affiliation{Physics Department, Hamilton College, Clinton NY 13323 USA}

\author{Bakir Husremovi\'c}
\affiliation{Physics Department, Hamilton College, Clinton NY 13323 USA}

\author{Samuel Marash}
\affiliation{Physics Department, Hamilton College, Clinton NY 13323 USA}

\author{Christopher Nadeau}
\affiliation{Booz Allen Hamilton, McLean, VA USA 22012}

\author{Man Nguyen}
\affiliation{Booz Allen Hamilton, McLean, VA USA 22012}

\author{Andrew M. Brownell}
\affiliation{Murray Associates, Utica, NY 13501 USA}

\author{Derrick Sica}
\affiliation{Murray Associates, Utica, NY 13501 USA}

\author{Michael Senatore}
\affiliation{Air Force Research Laboratory, Information Directorate, Rome NY 13441 USA}
\author{Samuel Schwab}
\affiliation{Air Force Research Laboratory, Information Directorate, Rome NY 13441 USA}
\author{Erin Sheridan}
\affiliation{Air Force Research Laboratory, Information Directorate, Rome NY 13441 USA}
\author{Matthew D. LaHaye}
\email{Matthew.LaHaye@us.af.mil}
\affiliation{Air Force Research Laboratory, Information Directorate, Rome NY 13441 USA}
\date{\today}
\begin{abstract}

We develop a compact transmon emitter/detector (TED) superconducting circuit and demonstrate its dual functionality as a single-photon source and detector. In our setup, photons emitted by a source TED are transmitted via a meter-long coaxial cable, routed through a circulator, and captured by a measurement TED. Both TED modules operate with nominally identical parameters, highlighting the flexibility of this novel architecture. Furthermore, we introduce an efficient microwave photon detection scheme tailored to the TED. Using this setup, we detect 60\% of the emitted Fock state photons and infer a 95\% detection efficiency at the input of the measurement TED, which we calibrate against coherent state measurements. The reset and photon emission/detection processes each require approximately $2\,\mu s$, yielding a minimum protocol duration of $4\,\mu s$ as constrained by our chosen TED parameters. Ultimately, the TED demonstrates a new use case for the recently developed double transmon coupler (DTC): a compact, drop-in, tunable, and transition-selective link between a coherent data transmon and a waveguide. Circuits like the TED will play a vital role in quantum information processing by facilitating unconditional fast reset, enabling microwave photon metrology, and serving as nascent quantum communication interfaces (QCIs). QCIs mediate entanglement distribution between quantum processing units (QPUs) within a cryostat or interface with microwave-to-optical transducers for long-range quantum networking.
\end{abstract}
\maketitle
\section{Introduction}

\begin{figure}
    \includegraphics[width=3.3in]{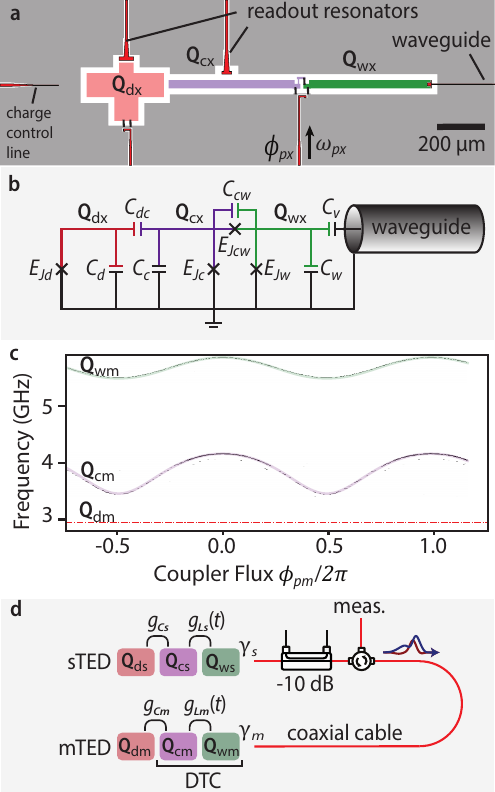}
    \caption{
    \textbf{(a)} CAD rendering of the transmon emitter/detector (TED) layout and
    \textbf{(b)} its corresponding circuit, indexed by $x\in\{s,m\}$ to distinguish between the source and measurement TEDs: sTED and mTED, respectively. The data transmon $Q_{dx}$ (red) is designed for long term storage, while $Q_{wx}$ (green) is strongly coupled to a waveguide. The coupling transmon $Q_{cx}$ (purple) and $Q_{wx}$ together form a double transmon coupler (DTC). An external flux line $\phi_{px}$ biases and sinusoidally drives the DTC to mediate interactions between $Q_{dx}$ and $Q_{wx}$.
    \textbf{(c)} Measured flux tunability of the mTED's $Q_{cm}$ (via two-tone spectroscopy) and $Q_{wm}$ (via coherent backscatter) transition frequencies with $\phi_{pm}$. The red dash-dot line is a guide to the eye indicating the $Q_{dm}$ transition, which varies by only $\sim 10$~MHz.
    \textbf{(d)} Schematic of the emission-detection setup. The sTED and mTED are connected via approximately 1 meter of coaxial cable, routed through a 10~dB directional coupler and a circulator. This configuration permits coherent backscatter measurements of $Q_{ws}$ and $Q_{wm}$, which lack an independent readout resonator. The transmon $Q_{dx}$ is capacitively coupled to $Q_{cx}$ with amplitude $g_{Cx}$, while $Q_{cx}$ and $Q_{wx}$ are inductively coupled with amplitude $g_{Lx}(t)$. The relaxation rate of $Q_{dx}$ into the waveguide is denoted by $\gamma_x$.
    }
    \label{fig:fig1}
\end{figure}

Networking physically separate quantum processing units (QPUs) is a compelling strategy for scaling quantum computers~\cite{Cirac1999, Kimble2008, Monroe2014, Jiang2007, Main2025} and realizing quantum advantage in distributed sensor networks~\cite{Zhao2021,Malia2022,Liu2021,Hainzer2024,Kim2024,Nichol2022}.
Superconducting circuits, which exhibit strong near-field interactions with one-dimensional waveguides, are well-positioned to integrate networking with scaled quantum processing.
To date, such circuits have realized the high fidelity creation~\cite{Houck2007, Besse2020, Miyamura2025, OSullivan2025, Wang2025, Kannan2023}, capture~\cite{Kurpiers2018, Grebel2024, Campagne-Ibarcq2018, Axline2018, Almanakly2025, Narla2016}, and detection~\cite{Chen2011, Koshino2013, Inomata2016, Kono2018, Besse2018, Lescanne2020, Balembois2024, Petrovnin2024, Opremcak2018, pankratov2025} of single microwave photons. These are essential functionalities of a quantum communication interface (QCI) architecture, analogous to classical serial network interfaces.
An effective QCI for a transmon-based architecture should ideally be compact, programmable in both transmitter and receiver modes, and support a broad frequency range. Crucially, it must provide sufficient isolation to prevent QPU decoherence while allowing for selective entanglement with the host QPU.

In this work, we demonstrate a compact, frequency-tunable, and transmon-compatible QCI: the transmon emitter/detector (TED). The TED mediates time-dependent interactions between a well-isolated, long-lived ``data'' transmon and the modes of a coplanar waveguide. The TED is the first device to utilize the recently developed double transmon coupler (DTC) architecture~\cite{Campbell2023, Kentaro2023, Li2024, kubo2024, Tiwari2026}, developed independently by Toshiba and the Air Force Research Laboratory (AFRL), to interface a data transmon with photonic waveguide modes. To showcase the versatility of the TED circuit, we prepare two devices with identical circuit parameters in a ``pitch-and-detect'' configuration. One device serves as a single microwave photon Fock source while the second functions as a microwave photon detector, indexed with $x=s$ and $x=m$, respectively. These chips are linked via a coaxial cable and a circulator, the latter of which enforces a continuum of modes and retrieves the outgoing waveform (Fig.~\ref{fig:fig1}a-d).

The TED circuit consists of the data transmon $Q_{dx}$ capacitively coupled to a DTC, which itself comprises two transmons ($Q_{cx}$ and $Q_{wx}$) linked by an inductive Josephson junction (Fig.~\ref{fig:fig1}a-b). This configuration effectively isolates $Q_{dx}$ from the waveguide environment. That environment is instead strongly coupled to $Q_{wx}$ on the opposite side of the DTC (See Appendix \ref{sect:device} for chip fabrication information). The frequency of $Q_{wx}$ therefore determines the energy of emitted or detected photons while its relaxation rate $\gamma_x$ to the waveguide sets the bandwidth.

By modulating the DTC flux, we hybridize $Q_{wx}$ and $Q_{dx}$ to operate the TED while keeping the middle ``coupling'' transmon $Q_{cx}$ far off resonance. Unlike previous applications of the DTC intended for high-fidelity two-qubit gates~\cite{Li2024, Tiwari2026} between data qubits, we flux drive the DTC at the difference frequency of $Q_{dx}$ and $Q_{wx}$ to establish a parametric exchange interaction between $Q_{dx}$ and the waveguide modes.

The TED offers several distinct advantages over existing QCI implementations. First, unlike cavity-based QCIs~\cite{Houck2007, Miyamura2025, Lescanne2020, Balembois2024, Kindel2016, Narla2016, Opremcak2018}, the TED is tunable. A static flux on the DTC tunes $Q_{wx}$ frequency over several hundred MHz, relaxing the stringent waveguide frequency matching requirements typically needed for entanglement distribution between heterogeneous chips/devices. Second, the frequency of $Q_{dx}$ is static and therefore compatible with the restrictive frequency-setting requirements of standard transmon QPUs. Finally, most tunable architectures require $Q_{dx}$ and $Q_{wx}$ to have frequencies within a couple hundred MHz of one another~\cite{Kannan2023, Almanakly2025, Forn-Diaz2017, Besse2020}. The TED circuit does not place restrictions on the frequency of $Q_{wx}$ relative to $Q_{dx}$: allowing simultaneous QPU and network interoperability.

In the following sections, we detail the TED's theory of operation, characterize its performance as a standalone source and detector, and finally demonstrate the end-to-end emission and detection of microwave Fock state photons. Simultaneously, we introduce a photon detection scheme suitable for the TED, perform an in-depth analysis of coherent reflective scattering (backscatter) measurements (including how to perform thermometry of the waveguide), and model the composite system and Fock state using SLH theory.

\section{Theory of operation}

The TED uses three transmons with specialized roles, indexed as $q\in\{d,c,w\}$ to denote the data, coupling, and waveguide transmons, respectively. We define the corresponding lowering operators for each respective transmon as $d_x$, $c_x$, and $w_x$. The internal connectivity of the TED is designed to provide static frequency and high isolation for the data transmon $Q_{dx}$ while allowing for tunable parametric coupling to the waveguide. $Q_{dx}$ and $Q_{cx}$ are capacitively coupled with magnitude $g_{Cx}$ (Fig.~\ref{fig:fig1}d). The coupling between $Q_{cx}$ and $Q_{wx}$ is modeled as inductive, $g_{Lx}$, as the static capacitive component plays a subtle role in the driven dynamics.

A dimensionless parameter $\phi_{px}(t) = \bar{\phi}_{px}+\tilde{\phi}_{px}\sin{(\omega_{px}t)}$ represents a flux modulation on the DTC. This yields an effective inductive coupling
\begin{align}
g_{Lx}(t)\sim \bar{g}_{Lx}-\tilde{g}_{Lx}\sin{(\omega_{px}t)}
\end{align}
where $\bar{g}_{Lx} = g_{L0x}\cos{(\bar{\phi}_{px})}$ and $\tilde{g}_{Lx} = \tilde{\phi}_{px}g_{L0x}\sin{(\bar{\phi}_{px})}$ (see Ref.~\cite{Campbell2023} for the exact functional form).

For a TED that is coherently driven through its waveguide (Appendix~\ref{sec:TEDH}) the system density matrix $\rho$ evolves according to the effective master equation $\dot\rho_x = -\mi [H_x, \rho_x]/\hbar + \gamma_x\mathcal{L} \!\left[ w_x \right] \rho_x$, where $\mathcal{L}$ is the Lindblad dissipator and $\gamma_x$ is the relaxation rate of $Q_{wx}$ into the waveguide. The Hamiltonian is defined as
\begin{align}
\frac{H_x}{\hbar} ={}&
\sum_{q=d,c,w}\omega_{qx}q_x^\dagger q_x+\nu_{qx} q_x^{\dagger 2}q_x^2\nonumber\\
&+\Big[
 g_{Cx} \left( d+d_x^{\dagger}\right)\left(c_x + c_x^{\dagger} \right)\nonumber\\ 
 &\qquad - g_{Lx}(t) \left(c-c_x^{\dagger}\right)\left(w_x - w_x^{\dagger} \right)\nonumber\\
&\qquad -i\Omega_x(t)\sin{(\omega_{\alpha x}t)} (w_x-w_x^{\dagger})\Big].
\label{eq:TEDH}
\end{align}
Here, $\nu_{qx}$ and $\omega_{qx}$ are the anharmonicity and frequency of each transmon, respectively. While the first line of Eq.~\eqref{eq:TEDH} gives the anharmonic oscillator level structure of the three transmons, the middle lines give the effective transmon-transmon interactions, and the last gives the response of $Q_w$ to a coherent drive with amplitude $\Omega_x(t)$ and frequency $\omega_{\alpha x}$ (used only for detector calibration).

When driving the DTC ($\tilde{\phi}_{px}(t)\ne 0$), the amplitudes $g_{Lx}$, $\omega_{qx}$, and neglected terms proportional to $q_x$ and $q_x^{\dagger}$ will oscillate. These oscillations produce the desired parametric driving term as well as frequency shifts that scale with $\tilde{g}_{Lx}$, and effective Rabi drives on each qubit, respectively. Therefore, we must carefully set the frequencies of $Q_{dx}$, $Q_{cx}$, and $Q_{wx}$ to avoid unwanted Rabi transitions.

To accurately estimate the lifetime of $Q_{dx}$, we avoid standard Purcell formulas which are rarely valid for superconducting circuits (Ref.~\cite{Yen2025}). Instead, we use a perturbative $RC$ time-constant approach~\cite{Esteve1986, Houck2008}, replacing the coaxial cable with a $50\,\Omega$ resistor to ground.
\begin{align}
\Gamma \approx \frac{\Re(Y)}{C_d+C_{dc}}\label{eq:admittancemain}
\end{align}
where $Y$ is the admittance to ground at the $Q_{dx}$ node (Fig.~\ref{fig:fig1}b). We write out $Y$ for $Q_{dx}$ in Appendix~\ref{sec:admittance}. 
We estimate the waveguide limits $Q_{dx}$ to a $T_1<13\,\textrm{ms}$ at $\bar{\phi}_{px}=0$, where coupling is maximal.

During operation, the DTC flux ($\phi_{px}$) is modulated at the difference frequency $\omega_{px}=|\omega_{wx}-\omega_{dx}|$.
A second time dependence is introduced by the coherent drive $\omega_{\alpha x} \sim\omega_{wx}$, used during measurement TED characterization. We apply two independent rotating-wave transformations and neglect fast rotating terms, to make Eq.~\eqref{eq:TEDH} stationary (see Appendices~\ref{sec:Hamiltonian_derivation}). In the new basis we adiabatically eliminate the far-off-resonance $Q_{cx}$ (see Appendix~\ref{sec:adiabatic_elimination}) to obtain an effective parametrically driven TED Hamiltonian
\begin{align}
    \frac{H_{\textrm{RWA}x}}{\hbar} = {}&
    \Big[(\delta_x-\delta_{px}) d_x^{\dagger}d_x 
    + \delta_x \, w_x^{\dagger}w_x 
    + \Omega_x(t)(w_x+w_x^{\dagger})/2
    \nonumber\\
    &+\nu_{dx} d_x^{\dagger 2} d_x^2 +\nu_{wx} w_x^{\dagger 2} w_x^2
    \nonumber\\
    &-ig_{px}(t) \left( d_x^{\dagger} w_x - d_x w_x^{\dagger} \right) \Big].
    \label{eq:TEDHeff}
\end{align}
where the qubit-drive detunings are $\delta_x = \omega_{wx}-\omega_{\alpha x}$ and $\delta_{px} = \omega_{wx} - \omega_{dx} - \omega_{px}$ and the effective coupling is $g_{px} = g_{Cx} \tilde{g}_{Lx}/2(\omega_{dx}-\omega_{cx})$. This $g_{px}(t)$ term mediates exchange transitions such as $|10\rangle\leftrightarrow |01\rangle$, hybridizing $Q_{dx}$ and $Q_{wx}$ to facilitate state reset or photon emission/detection. The time dependent Schrodinger equation acts on the two-transmon state $|n_{dx} n_{wx}\rangle$, where $n_{dx}$ and $n_{wx}$ represent the excitation level of $Q_{dx}$ and $Q_{wx}$, respectively.

\subsection{Device  characterization}
\label{sec:characterization}

\begin{figure}
    \includegraphics[width=3.3in]{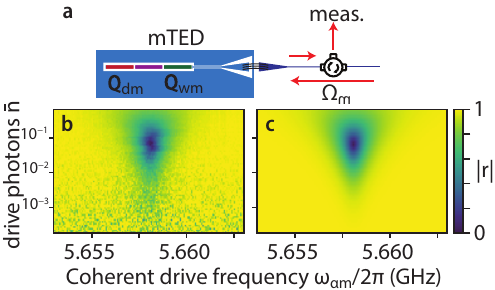}
    \caption{\textbf{(a)} Schematic layout of the mTED. A coherent drive tone from a vector network analyzer is applied to the waveguide-coupled transmon $Q_{wm}$ while the parametric coupling is disabled ($g_{pm}=0$). \textbf{(b)} Measured and \textbf{(c)} simulated coherent backscatter response, represented by the normalized complex reflection amplitude $r$ from $Q_{wm}$. Fitting the theoretical simulation to the experimental data calibrates the coherent drive power and extracts key system parameters: the $Q_{wm}$ resonance frequency, the relaxation rate $\gamma_m$, and the waveguide thermal population.}
    \label{fig:calibration}
\end{figure}

Fitting simulation to a measurement of the normalized amplitude of the coherent state (backscatter) reflected from the waveguide transmon $Q_w$ (Fig.~\ref{fig:calibration}a-c) provides a robust way to extract $\gamma_x$, $\omega_{wx}$, and the thermal population of the waveguide. The response of $Q_w$ to a coherent drive is inherently nonlinear and power-dependent, categorized into three distinct scattering regimes: elastic, inelastic, and saturated.

In the elastic regime, the incident drive power is sufficiently low that $Q_w$ mimics a linear harmonic oscillator, as the excitation probability to the second excited state remains negligible. In this limit, the qubit-emitted field interferes with the reflected drive to produce a return signal with equal magnitude and opposite sign to the non-interacting reflected drive. However, non-zero thermal occupation $n_{th}$ in the waveguide occasionally excites $Q_{wx}$. Since the transitions to the state above and below any excited state of $Q_{wx}$ are dissimilar, there is an  observable nonlinearity even for arbitrarily weak coherent drives, suppressing the coherent emission amplitude from $Q_{wx}$.
As the drive power increases into the intermediate regime, the coherent emission continues to decrease. Our model (Fig.~\ref{fig:calibration}c) predicts a specific power $P = \hbar \omega_{wx} \gamma_x / 16$ at which the field emitted by $Q_{wx}$ and the reflected drive destructively interfere, resulting in a characteristic dip in the return signal on a vector network analyzer. In the presence of thermal photons this cancellation occurs at a weaker power. Finally, in the saturated regime, the qubit becomes a negligible perturbation to the high-power drive, and the scattering response converges to that of a waveguide with no qubit.

By fitting across all three regimes, we extract the resonance frequency $\omega_{wx}/2\pi=5.65811(1)$~GHz, the decay rate $\gamma_m =11.2(1)\times 10^6\,\mathrm{s^{-1}}$, and a thermal occupation $n_{th}=0.015(3)$. This thermal population is consistent with the occupation of $Q_{dm}$ observed after the reset procedure detailed in section~\ref{sec:source}.

The anharmonicity of $Q_{wx}$ is determined by monitoring the reflected response in the elastic or intermediate power regime while sweeping a second, independent frequency. When the frequency of the second drive is resonant with the $|01\rangle\leftrightarrow |02\rangle$ transition, the response of the primary probe is suppressed~\cite{Abdumalikov2010}. From this, we find the anharmonicity of $\nu_{wx}$ is $-169(2)$~MHz.

The transition frequencies and anharmonicities (Table~\ref{tab:params} in Appendix~\ref{sect:device}) for $Q_{cx}$ and $Q_{dx}$ were characterized via standard two-tone spectroscopy~\cite{Krantz_apr2019} through their respective readout resonators.
At its upper flux sweet spot, $Q_{dx}$ exhibits a lifetime $T_1 = 81(35)\,\mathrm{\mu s}$ and Hahn echo coherence $T_2 = 41(6)\,\mathrm{\mu s}$, both of which remain independent of the DTC static flux bias $\bar\phi_{px}$. Crucially, the lifetime and coherence are unaffected by the application of the parametric drive ($g_{px} = 0.584(6)\gamma_m$) during TED operation, provided no photons are actively incident on $Q_{wx}$.

Finally, by mapping the DTC transmon frequencies as a function of DTC flux and fitting to the model in Ref.~\cite{Campbell2023}, we infer an inductive coupling $g_{L0}=(0.30(1)\text{ GHz})\cos{(\bar\phi_p)}$ at $\bar\phi_{px} = 0$. Finite element simulations (ANSYS~Q3D) indicate a capacitive coupling $g_{Cx}/2\pi \approx 70$~MHz between $Q_{dx}$ and $Q_{cx}$.

\section{Source and detector operation}

In the experiments detailed in sections~\ref{sec:source}-\ref{sec:pitch-detect}, the source TED (sTED) and measurement TED (mTED) are treated as a composite two-TED system (Fig.~\ref{fig:fig1}d). We denote the system states as $\ket{\textrm{sTED}}\otimes \ket{\textrm{mTED}}$, using the shorthand $\ket{\textrm{sTED}}\otimes \ket{\cdot \cdot}$ or $\ket{\cdot \cdot}\otimes\ket{\textrm{mTED}}$ when the state of the mTED or sTED can be traced out without influencing the discussion, respectively.

\subsection{Source operation}
\label{sec:source}

\begin{figure}
    \includegraphics[width=3.3in]{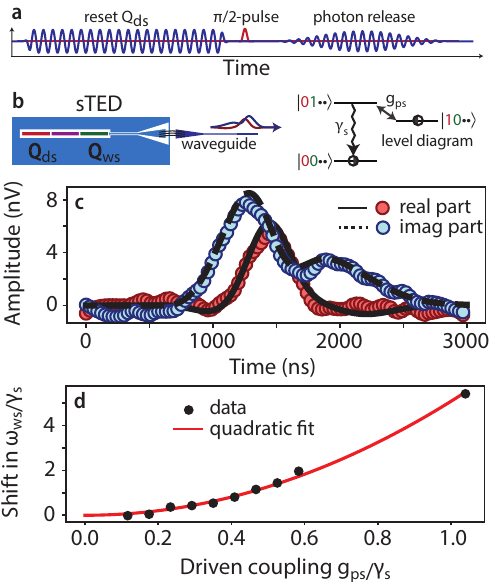}
    \caption{\textbf{(a)} Pulse timing diagram for parametric coupling and qubit gates. An initial DTC flux drive resonant with the $|10\cdot\cdot\rangle \leftrightarrow |01\cdot\cdot\rangle$ transition resets the residual thermal population of $Q_{ds}$. Next, a $\pi/2$-pulse prepares $Q_{ds}$ in the superposition state $(\ket{10\cdot\cdot}+\ket{00\cdot\cdot})/\sqrt{2}$. A shaped flux drive on the reset transition subsequently releases the photon state.
    \textbf{(b)} Left: layout of sTED, where a photon emits into a $50\,\Omega$ waveguide. Right: level diagram illustrating $Q_{ds}$ prepared in a superposition state (indicated by half filled circles). Here, $g_{ps}$ represents the parametric drive amplitude used to release the state from $Q_{ds}$, and $\gamma_s$ is the relaxation rate of $Q_{ws}$ into the waveguide.
    \textbf{(c)} Time-domain complex amplitude of the emitted photon superposition state. Markers indicate the averaged amplified signal over $10^4$ repetitions (error bars are smaller than the markers); these were vertically rescaled to match the theory fit (lines). The complex wavefunction dynamics arise from a shift in $\omega_{ws}$ dependent on $g_{ps}/\gamma_s$. These shifts can be passively eliminated (see Appendix~\ref{sec:acStark})
    \textbf{(d)} Measured frequency shift of $\omega_{ws}$ as a function of $g_{ps}/\gamma_s$. The solid line is a quadratic fit to the data.
    \label{fig:fig2}}
\end{figure}

We use the pulse sequence illustrated in Fig.~\ref{fig:fig2}a-b to generate a $(\ket{10\cdot\cdot}+\ket{00\cdot\cdot})/\sqrt{2}$ superposition state in $Q_{ds}$ and subsequently emit a Fock-vacuum superposition state into the waveguide.

Prior to state preparation, we apply a $2\,\mu s$ ``reset'' parametric drive $g_{ps}(t)$ to hybridize $Q_{ds}$ with $Q_{ws}$. Without this step, we observe a residual thermal population of $12\,\%$ for $|10\cdot\cdot\rangle$. The reset pulse brings $Q_{ds}$ into equilibrium with the waveguide's mode occupation at $\omega_{ws}$, reducing its population to $1.5\,\%$ and ensuring consistent state initialization.
We set the emission frequency $\omega_{ws}$ via the DTC's static flux bias $\bar\phi_{ps}$. An initial $\pi/2$ pulse is applied on the $Q_{ds}$ charge line, followed by a modulation of the parametric coupling $g_{ps}(t)$ at frequency $\omega_{ps}=|\omega_{ws}-\omega_{ds}|$. This modulation hybridizes $Q_{ds}$ with $Q_{ws}$, which decay into the waveguide at rate $\gamma_s$ to emit the superposition state.

Unlike a Fock state, a Fock-vacuum superposition wavefunction has a non-zero time averaged amplitude and phase. We show the average of ten~thousand measurements post-amplification in Fig.~\ref{fig:fig2}c using the protocol described above. The time and frequency dependence of an equivalent Fock state wavefunction, produced when $Q_{ds}$ is instead fully excited during initialization, can be inferred from this averaged measurement~\cite{Eichler2011, Eichler2012, Kannan2023}.

We do not carefully shape the emitted wavefunction as other works have~\cite{Miyamura2025,Almanakly2025} because detection only requires that the photon wavefunction have a frequency profile that falls within the mTED's bandwidth (see Sec.~\ref{sec:pitch-detect}).
To achieve this, we modulated $g_{ps}\propto \tilde{\phi}_{ps}$ with a cosine shape
\begin{equation}
\tilde{\phi}_{ps}(t) = \tilde{\phi}_{ps}\cos^2\!\left( \frac{\pi (t-t_0)}{T} \right)\label{eq:parametric_drive}
\end{equation}
for $t-t_0\in (-T/2,T/2)$ (as illustrated in Fig.~\ref{fig:fig2}a). We simulated and measured that, with duration $T = 2~\mu\text{s}$, this shape and parametric drive amplitude is statistically likely to release a photon into the waveguide with a narrow frequency profile.

We measured a quadratic shift of $\omega_{ws}$ with applied $g_{ps}$ (Fig.~\ref{fig:fig2}d). Note that the phase of the wavefunction amplitude has a linear slope with time when there is no shift. Incorporating this frequency shift into Eq.~\ref{eq:TEDHeff} and letting $g_{ps}$ vary as a free parameter we obtain $g_{ps}(t) = 0.472(2)\gamma_s$ at $t = t_0$. There are two roughly co-equal origins to the shift. First, the off-resonant interaction of the modulating inductive coupling $\tilde{g}_{Ls}$ with $Q_{cs}$ induces an ac~Stark shift of $\omega_{ws}$ equivalent to $\tilde{g}_{Ls}(t)^2/4(\omega_{ds}-\omega_{cs})$. Second, the dispersion of $\omega_{ws}(\phi_{ps})$ has some curvature about most bias points $\bar{\phi}_{ps}$, resulting in some rectification. We subsequently fabricated and measured a circuit with substantially larger $g_{Cs}$ and $E_{Jcws}$ that had a shift much smaller than $\gamma_s$ at $g_{ps}\approx \gamma_s/2$. In Appendix~\ref{sec:acStark} we discuss some of the nuances involved in choosing these circuit parameters.

\subsection{Detector operation}
\label{sec:detector}

\begin{figure}
    \includegraphics[width=3.3in]{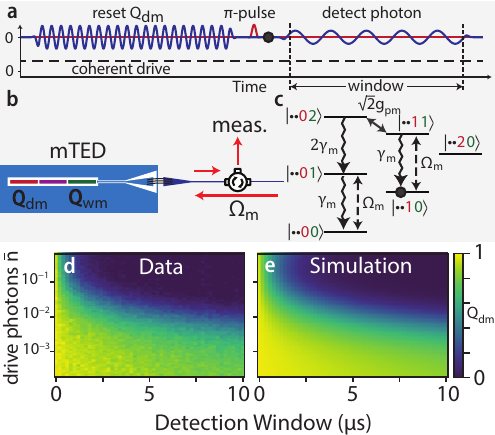}
    \caption{\textbf{(a)} Pulse timing diagram for the mTED under continuous external coherent driving. The mTED is initialized using a reset pulse on the $|\cdot\cdot10\rangle\leftrightarrow|\cdot\cdot01\rangle$ transition, followed by a $\pi$-pulse to excite $Q_{dm}$ to the $\ket{\cdot\cdot10}$ state. Detection is activated by parametrically driving the $|\cdot\cdot11\rangle\leftrightarrow|\cdot\cdot02\rangle$ transition.
    \textbf{(b)} Layout of the mTED, where a coherent drive is applied to $Q_{wm}$.
    \textbf{(c)} Energy level diagram indicating the driven transitions and relaxation pathways. The parametric drive amplitude is $g_p$, and $Q_{wm}$ relaxes into the waveguide at a rate $\gamma_m$.
    \textbf{(d)} Measured and \textbf{(e)} simulated population of $Q_{dm}$ as a function of detection window duration and coherent drive power expressed in drive photons $P=\gamma_m\bar{n}$ (population of $Q_{dm}$ in response to this drive will differ based on architecture and excitation level).}
    \label{fig:fig3}
\end{figure}

To contextualize the operation of the mTED, we first distinguish between microwave photon detection and microwave photon capture. A capture protocol~\cite{Kurpiers2018, Grebel2024, Campagne-Ibarcq2018, Axline2018, Almanakly2025, Narla2016} swaps the quantum state (up to a single photon) of a single waveguide mode with $Q_{dm}$, allowing for full quantum state reconstruction via tomography. Alternatively, microwave photon detection~\cite{Chen2011, Koshino2013, Kono2018, Besse2018, Lescanne2020, Balembois2024, Petrovnin2024, Opremcak2018} integrates over the waveguide within the defined detection window and bandwidth $\gamma_m$. If a photon is present, $Q_{dm}$ ``latches'' to its ground state. While capture protocols allow one to infer the phase of a Fock-vacuum superposition like $(|1\rangle-|0\rangle)/\sqrt{2}$, detection protocols only relay whether a Fock state was incident on the detector. At optical frequencies, physical implementations of photon detectors include photomultiplier tubes and superconducting nanowire single photon detectors. 

Detection protocols often transfer the incoming photon to a distinguishable set of empty photonic modes (Refs.~\cite{Kono2018, Besse2018} are notable exceptions), conditioned on a change of a qubit's state. While prior implementations use a separate destination waveguide~\cite{Inomata2016,Koshino2013,Lescanne2020, Balembois2024}, our protocol releases the photons back into the same waveguide at a new frequency. 

We initiate the detection protocol by preparing the mTED in the $|\cdot\cdot10\rangle$ state (black dot in Fig.~\ref{fig:fig3}a) via a $\pi$-pulse on $Q_{dm}$. As illustrated in the schematic and level diagram (Fig.~\ref{fig:fig3}b-c), an incoming photon at frequency $\omega_{wm}$ then triggers the transition $\ket{\cdot\cdot10}\rightarrow\ket{\cdot\cdot11}$. If we modulate the DTC flux at $\omega_{pm} = |\omega_{wm}-\omega_{dm}|+\nu_{wm}$ while the photon is incident, we parametrically drive the $\ket{\cdot\cdot11}\leftrightarrow \ket{\cdot\cdot02}$ transition where $\ket{\cdot\cdot02}$ then irreversibly relaxes to its ground state. Consequently, $Q_{wm}$ only reaches the intermediate $\ket{\cdot\cdot02}$ state if a photon arrives along the waveguide while the parametric drive is active.

The efficiency can approach unity, particularly when $\gamma_m/2\pi$ significantly exceeds the frequency extent of the incident photon's wavefunction. The absorbed photon is re-emitted as a correlated photon pair $\omega_{wm}$ and $\omega_{wm}+\nu_{wm}$~\cite{Gasparinetti2017} while the mTED latches to $|\cdot\cdot 00\rangle$ (See appendix~\ref{sec:avgmeas}).

Relaxation of $Q_{dm}$ during photon detection and readout corresponds to a false detection event (or dark count) using this protocol. We can expect $\sim 16\,\%$ false detection events with an $81(35)\,\mathrm{\mu s}$ lifetime for $Q_{dm}$, a $10\,\mathrm{\mu s}$ detection window, and $4\,\mathrm{\mu s}$ readout duration. False detection events are undesirable for some quantum networking applications, such as entanglement heralding, via DLCZ~\cite{Duan2001} or Barrett-Kok~\cite{Barrett2005} protocols, within which false detector click events contribute to infidelity in the generated distributed Bell state.

We compared the behavior of the mTED to our model (Eq.~\ref{eq:TEDHeff} and Appendix~\ref{sec:scatQw}) with the amplitude of $g_{pm}$ as a free parameter. Using the setup described in Fig.~\ref{fig:fig3}a-c, we monitored the population of $Q_{dm}$ as a function of coherent drive power $\hbar\omega_{wm}\gamma_m \bar n$ applied along the waveguide at frequency $\omega_{wm}$ and the detection window (when $g_{pm}\ne 0$). Fitting simulation to data in Fig.~\ref{fig:fig3}d-e gives $g_{pm} = 0.259(3)\gamma_m$.
Detection efficiency can reach near unity for any $g_{pm}\ge\gamma_m/2$. The choice of $g_{pm}=\gamma_m/2$ maximizes the detection bandwidth at $\omega_{wm}$. For $g_{pm}>\gamma_m/2$ there are two optimal frequencies of detection, separated by $2g_{pm}$. The procedure for calibrating the coherent drive power at the input of the mTED is described in section~\ref{sec:characterization}.

\subsection{Pitch-detect demonstration}
\label{sec:pitch-detect}
We use a pitch-and-detect demonstration to explore Fock state detection. Here, the sTED emits a single microwave Fock state that is subsequently detected by the mTED. The two TED devices are linked via approximately one meter of non-superconducting coaxial cable, a directional coupler, and a circulator, as illustrated in Figs.~\ref{fig:fig1}d and \ref{fig:fig4}b. The circulator enforces a continuum of modes, routing the itinerant photon from the sTED to the mTED while simultaneously directing any field emitted or reflected by the mTED toward the measurement chain.

We formulate a theoretical description of the coupled sTED-mTED system interacting via a circulator using the SLH formalism~\cite{SLH1, Gough_2008, Combes_2017}. Incorporating transmission losses through the cabling and circulator via a phenomenological loss parameter $\eta \in [0,1]$, the evolution of the joint density matrix $\rho$ is governed by the effective master equation (see Appendix~\ref{sec:SLH}):
\begin{align}
\dot\rho = {}&-\frac{\mi}{\hbar} [H, \rho] + \gamma_s \mathcal{L} \!\left[ w_s \right] \rho + \gamma_m \mathcal{L} \!\left[ w_m \right] \rho \nonumber
\\
&+ \sqrt{(1-\eta^2) \gamma_s \gamma_m} \Bigl( \mathcal{L}_2 \!\left[ w_s, w_m \right] \rho \nonumber
\\
&+ \mathcal{L}_2 \!\left[ w_m, w_s \right] \rho \Bigr)
\label{eq:rhoSLH}
\end{align}
where $\mathcal{L}_2[A, B] \rho = A \rho B^\dagger - \frac{1}{2} \{ A^\dagger B, \rho\}$ is a generalized Lindblad dissipator and the Hamiltonian is given by
\begin{align}
H ={}& H_s + H_m + \frac{\mi}{2} \sqrt{(1-\eta^2) \gamma_s \gamma_m} \left( w^{\dagger}_s w_m - w_s w^{\dagger}_m \right).
\label{eq:HSLH}
\end{align}
Note that in Eq.~\ref{eq:HSLH}, we have omitted terms accounting for the dynamical phase accumulated by the photon during its propagation through the coaxial cabling. This simplification is justified because the detection protocol irreversibly entangles the state of $Q_{dm}$ with additional photons emitted into the continuum. Tracing over these unmeasured photonic degrees of freedom yields an effective bit flip to ground on $Q_{dm}$ that is insensitive to both the phase of an equatorial projection of $Q_{dm}$ and the phase of a Fock-vacuum superposition in the waveguide incident on the mTED. Derivations of Eq.~\eqref{eq:HSLH}—phase evolution terms included—and the corresponding collapse operators in Eq.~\eqref{eq:rhoSLH} are provided in Appendix~\ref{sec:SLH}

\section{Results}

\begin{figure}\includegraphics[width=3.3in]{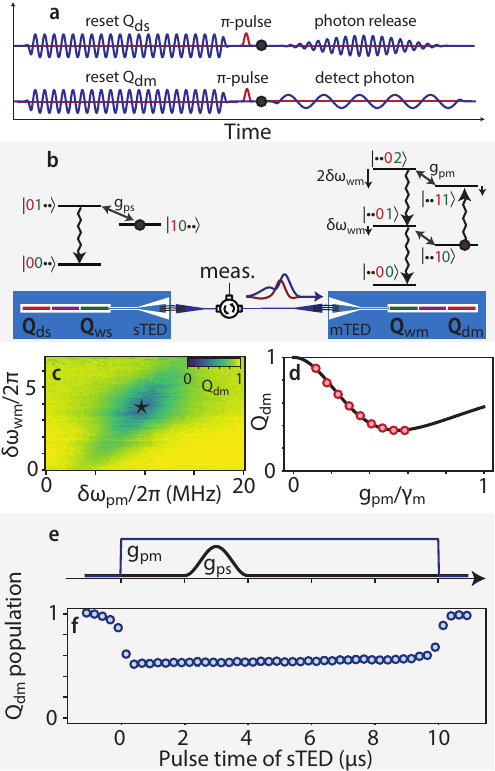}
    \caption{\textbf{(a)} Pulse protocols for the parametric couplings $g_{ps}$ (top) and $g_{pm}$ (bottom), alongside the corresponding data qubit operations (red traces) on $Q_{ds}$ and $Q_{dm}$. \textbf{(b)} Energy level diagrams indicating the driven transitions. Left: Activating $g_{ps}$ drives an excitation from $Q_{ds}$ into $Q_{ws}$, which rapidly decays to emit a photon into the waveguide. A circulator routes the itinerant photon and eliminates resonant cable modes. Right: when the photon arrives at the mTED, the population of $Q_{dm}$ goes to zero. \textbf{(c)} Normalized population of $Q_{dm}$ (inversely proportional to the photon detection probability) plotted against frequency excursions $\delta\omega_{wm}$ and $\delta\omega_{pm}$ from $\omega_{wm}$ and $\omega_{wm}-\omega_{dm}-\nu_{wm}$. These excursions are controlled by varying $\bar{\phi}_p$ and $\omega_p$, respectively. The star indicates the optimal ($\delta\omega_{wm}$,$\delta\omega_{pm}$) for maximum detection probability. \textbf{(d)} Measured $Q_{dm}$ population as a function of $g_{pm}/\gamma_m$ at the optimal point from panel~(c). The solid line represents the theoretical prediction using the network loss $\eta$ as a free parameter. \textbf{(e)} Timing schematic showing the photon release pulse $g_{ps}$ (here, at $3\,\mu s$) relative to the active detection window $g_{pm}$. \textbf{(f)} Detection fidelity plotted as a function of the photon release pulse time, demonstrating that detection probability is largely unchanged by the arrival time of the photon within the window.}
    \label{fig:fig4}
\end{figure}

We implemented the pitch-and-detect configuration detailed in Section~\ref{sec:pitch-detect}. The experimental protocol, depicted in Fig.~\ref{fig:fig4}a, begins with the simultaneous excitation of $Q_{ds}$ and $Q_{dm}$, initializing the system in the state $|1010\rangle$. The sTED then emits a Fock state photon during the active detection window of the mTED. Upon a successful detection event, the mTED emits correlated photons into its waveguide, which are subsequently routed to the output measurement chain as illustrated in Fig.~\ref{fig:fig4}b. Using the measured photon wavefunction from the sTED (section~\ref{sec:source}) and numerical simulations of the mTED dynamics during a detection event, we estimate that approximately $5\%$ of the incident photon wavefuction is clipped by the mTED's bandwidth.

We obtain the average excited-state population of $Q_{dm}$ as a function of both $\phi_{pm}$ (which tunes $\omega_{wm}$) and the parametric drive frequency $\omega_{pm}$ applied to the DTC (Fig.~\ref{fig:fig4}c).
Co-optimizing these parameters establishes the optimal resonance condition (indicated by the black star) for absorbing the incident sTED photon. Using a $2\,\mathrm{\mu s}$ detection window followed by a $4\,\mathrm{\mu s}$ readout, we observe a baseline $Q_{dm}$ population of $92\,\%$ in the absence of incident photons from the sTED, corresponding to an $8\,\%$ dark count  rate that is consistent with the $T_1$ relaxation of $Q_{dm}$. In Figs.~\ref{fig:fig4}c-d,~and~f the $Q_{dm}$ population is normalized to a [0,1] range and is inversely proportional to the probability of detecting a photon. Numerical modeling indicates that for the optimal drive amplitude $g_{pm}=\gamma_m/2$, the intrinsic detection fidelity of the mTED reaches $95\,\%$. Comparing this baseline fidelity to the measured $Q_{dm}$ population, we infer a $35\%$ photon loss within the intervening cabling, which consists of approximately one meter of non-superconducting coaxial cable and the routing circulator. 

To highlight the classical behavior of the mTED, Figs.~\ref{fig:fig4}e-f show that varying the arrival time of the sTED photon within the mTED's $10~\mu \text{s}$ detection window has minimal effect on the detection probability. The slight slope in Fig.~\ref{fig:fig4}f at later arrival times is attributable to the $T_1$ relaxation of $Q_{ds}$ occurring between its initial excitation and the delayed photon release. The reduction in the peak detection probability observed in Fig.~\ref{fig:fig4}f relative to Fig.~\ref{fig:fig4}c-d results from extending the detection window from $2~\mathrm{\mu s}$ to $10~\mathrm{\mu s}$. The additional accumulation of $Q_{dm}$ relaxation events over the $10~\mathrm{\mu s}$ detection window increases the dark count rate to $16\,\%$.

\section{Conclusions}

Inserting a double transmon coupler (DTC) between a data transmon $Q_d$ and a networking waveguide yields a modular, drop-in architecture for microwave photon emission and detection that preserves the intrinsic lifetime and coherence of $Q_d$. Crucially, this design decouples the resonance frequency of $Q_d$ from that of the waveguide-coupled transmon ($Q_w$), accommodating the strict frequency-setting requirements of transmon quantum processing units (QPUs). The broad $\sim 300$~MHz flux-tunability of $Q_w$ significantly eases spectral matching for distributing entanglement between dissimilar QPU modules. Similarly, microwave to optical transducers exhibit considerable variability in their operational resonance frequencies in the microwave domain, hence the frequency mobility of $Q_w$ provides a mechanism for reliably interfacing superconducting QPUs with optical quantum networks.

Additionally, using the transition selectivity demonstrated during mTED detection, the TED architecture functions effectively as a transition-selective quantum communications interface (QCI). This capability is useful for photon-circuit entanglement and is essential to implementing type-II entanglement heralding protocols~\cite{Luo2009} with a single data qubit. Ultimately, in-situ reconfigurability is the defining advantage of the TED, allowing the flexible management of the QCI between a superconducting QPU and its networking components, including the ability to switch between source and detector modes of operation.

\begin{acknowledgments}
DLC, SM, MA, AM, VH, and ML gratefully acknowledge support from AFOSR under LRIR 22RICOR003. We’d like to thank Yuvraj Mohan and the Rigetti Quantum Foundry Services team for design support and device fabrication. The authors also gratefully acknowledge support from MIT Lincoln Laboratory and IARPA LogiQ program for the traveling wave parametric amplifiers (TWPAs) used in the measurement chains for this work. VH acknowledges support from Hamilton College. Any opinions, findings, and conclusions or recommendations expressed in this article are those of the authors and do not necessarily reflect the views of the Air Force Research Laboratory (AFRL). Approved for Public Release; Distribution Unlimited: PA\# AFRL-2026-2067.
\end{acknowledgments}
\appendix
\appendix

\section{The TED device}

    \label{sect:device}

The sTED and mTED devices were fabricated on 6.25~mm $\times$ 6.25~mm high-resistivity silicon chips using Rigetti foundry services. The circuit ground planes are connected via indium bump bonds to a cap chip in a flip-chip configuration. We measured all our flip-bonded chips to have crosstalk less than $1\%$ between flux lines.

\begin{figure}
    \includegraphics[width=3.3in]{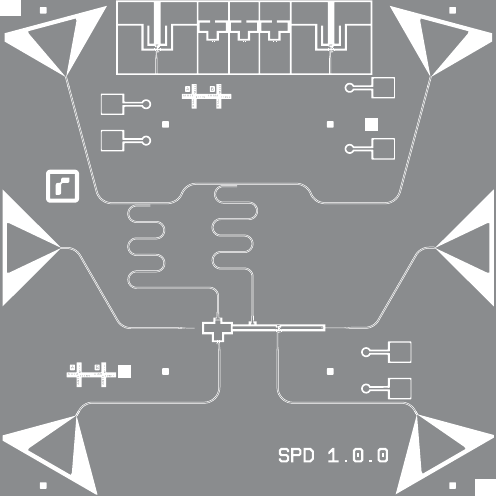}
    \caption{Full CAD layout of the fabricated device under test. A cap chip, attached via indium bump bonds to the chip shown here, connects the various ground planes.
    }
    \label{fig:layout}
\end{figure}

The TED device layout is shown in Fig.~\ref{fig:layout}
and parameters are shown in Table~\ref{tab:params}. 

\begin{table}[htp]
    \begin{tabular}{ll} 
    Parameter & Value\\
    \hline
    frequencies & \\
    $\omega_{ds}/2\pi$ & $3.155$~GHz\\
    $\omega_{dm}/2\pi$ & $2.95$~GHz\\
    $\omega_{cx}/2\pi$ & $3.87(1)$~GHz\\
    $\omega_{wx}/2\pi$ & $5.65811(1)$~GHz\\
    \hline
    coupling & \\
    $g_{Cx}/2\pi$ & $\sim0.07$~GHz\\
    $g_{Lx}(\Phi_p=0)/2\pi$ & $0.30(1)$~GHz\\
    \hline
    anharmonicities &  \\
    $\nu_{dx}/2\pi$ & $-0.174(6)$~GHz\\
    $\nu_{cx}/2\pi$ & $-0.169(2)$~GHz\\
    $\nu_{wx}/2\pi$ & $-0.169(2)$~GHz\\
    \hline
    decay rates & \\
    $\gamma_x$ & $11.2(1) \times 10^6 \,\mathrm{s}^{-1}$\\
    \hline
    drive frequencies & \\
    $\omega_{ps}$ & 2.5028~GHz\\
    $\omega_{pm}$ (reset) & 2.541~GHz\\
    $\omega_{pm}$ (detection) & 2.372~GHz\\
    \hline
    Junction energies & \\
    $E_{Jd}/h$ & 8.7 GHz\\
    $E_{Jc}/h$ & 13 GHz\\
    $E_{Jw}/h$ & 26 GHz\\
    $E_{Jcw}/h$ & 2.2 GHz\\
    \hline
    capacitances & \\
    $C_{d}$ & 121 fF\\
    $C_{c}$ & 112 fF\\
    $C_{w}$ & 110 fF\\
    $C_{dc}$ & 3.8 fF\\
    $C_{cw}$ & $\sim 7$~fF\\
    $C_{v}$ & 4.5 fF\\
    \hline
    $M_d$ & $1$~pH\\
    $M_p$ & $3$~pH
    \end{tabular}
    \caption{Table of parameters for the mTED and sTED. Indices $x\in\{s,m\}$ distinguish between the sTED and mTED, respectively. $M_d$ and $M_p$ are the mutual inductances of the respective flux lines for $Q_{dx}$ and the DTC of each TED.
    }
    \label{tab:params}   
\end{table}

\begin{figure}
    \includegraphics[width=3.3in]{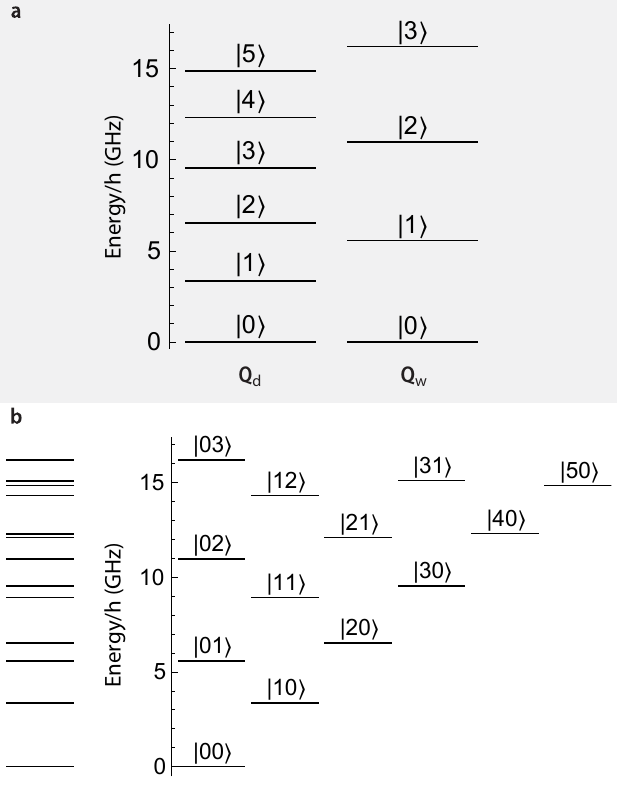}
    \caption{\textbf{(a)} Isolated energy levels for the data transmon $Q_d$ and the waveguide coupled transmon $Q_w$. \textbf{(b)} Composite energy levels of a TED, denoted by $|n_d\, n_w \rangle$. The diagram is organized such that each step to the right increments $n_d$ by one. The levels are overlaid in the left subpanel.
    }
    \label{fig:energylevels}
\end{figure}

The $Q_d$ and $Q_w$ transmon energy levels, calculated with Qutip~\cite{qutip5} and verified with Scqubits~\cite{scqubits2021}, are shown side-by-side in Fig.~\ref{fig:energylevels}a. In Fig.~\ref{fig:energylevels}b we organize the states $\ket{n_d \, n_w}$ so that each step to the right increments the excitation level of $Q_d$. Stepping vertically up or down corresponds to the absorption or emission of a photon at $Q_w$. In this construction, a long-lived state, such as an excited state of $Q_d$, does not have another state beneath it.

\section{TED Hamiltonian}
\label{sec:TEDH}

We define the charge and flux operators on transmon $q\in\{d,c,w\}$ of a single TED (hence we drop index $x\in\{s,m\}$) as $\hat{\mathcal{Q}}_q$ and $\varphi_0\hat\phi_q$, respectively, where $\varphi_0=\Phi_0/2\pi$ and $\Phi_0$ is the magnetic flux quantum. We use hats to help distinguish the charge and flux operators from the transmon and parametric external flux labels in the main text. Junction energies and capacitances have the circuit configuration shown in Fig.~\ref{fig:fig1}b. The TED Hamiltonian then takes the form
\begin{align}
H =& \boldsymbol{\mathcal{Q}}^T \check C^{-1} \boldsymbol{\mathcal{Q}}\label{eq:charge}\\ &- E_{Jd}\cos{(\hat\phi_d)} - E_{Jc}\cos{(\hat\phi_c+\phi_{pc})}\nonumber\\ &-E_{Jw}\cos{(\hat\phi_w+\phi_{pw})}\nonumber\\
&-E_{Jcw}\cos{(\hat\phi_w-\hat\phi_c+\phi_{pcw})}\nonumber\\
\end{align}
where
\begin{align}
\check{C} =& \begin{pmatrix} C_d + C_{dc} & -C_{dc} & 0\\
-C_{dc} & C_c+C_{dc}+C_{cw} & -C_{cw}\\
0 & -C_{cw} & C_w+C_{cw}+C_v
\end{pmatrix}\\
\boldsymbol{\mathcal{Q}}^T =& \begin{pmatrix}\hat{\mathcal{Q}}_d &\hat{\mathcal{Q}}_c &\hat{\mathcal{Q}}_w \end{pmatrix}\\
\phi_p =& \phi_{pc}+\phi_{pcw}+\phi_{pw}\\
\frac{\phi_{pc}}{\phi_p} =& \frac{C_{cw}C_w}{\textrm{det}(\check C)},\,\,\, \frac{\phi_{pcw}}{\phi_p} = \frac{C_{c}C_w}{\textrm{det}(\check C)},\,\,\,\frac{\phi_{pw}}{\phi_p} = \frac{-C_{cw}C_c}{\textrm{det}(\check C)}.\nonumber
\end{align}
Modulating the flux across an inductor also induces a time dependent charge across the capacitor in parallel with the inductor~\cite{You2019}. Our choice of the external flux $\{\varphi_0\phi_{pc},\,\varphi_0\phi_{pcw},\,\varphi_0\phi_{pw}\}$ dropped across each inductive branch cancels these induced charges. Other distributions of branch fluxes in Eq.~\ref{eq:charge} would yield a correct result if the appropriate time-dependent induced charge terms were included.

\subsection{Harmonic oscillator basis}

We convert to the harmonic oscillator basis to isolate terms of interest from Eq.~\ref{eq:charge}. Using index $q\in\{d, c, w\}$ we may make a substitution for the charge and flux operators
\begin{align}
\hat{\mathcal{Q}}_q = \sqrt{\frac{\hbar}{2 Z_q}}(a_q+a^{\dagger}_q)\\
\hat\phi_q = \frac{-i}{\varphi_0}\sqrt{\frac{\hbar Z_q}{2}}(a_q-a^{\dagger}_q)\\
[\varphi_0\hat\phi_q,\,\hat Q_q] = i\hbar
\end{align}
where a poor choice of $Z_q$ can still yield accurate numerical results if enough harmonic oscillator levels are included in a simulation. Using the spanning tree method~\cite{Vool_2017} to obtain $\check C$ and $\check L^{-1}$, we choose $Z_q = \sqrt{\check{C}^{-1}_{q,q}/\check{L}^{-1}_{q,q}}$ as the effective qubit impedance where $\check{C}^{-1}_{q,q}$ and $\check{L}^{-1}_{q,q}$ are diagonal elements of their respective matrices. A procedure for defining an inverse linear inductance for a junction element biased with an external flux is to first substitute $E_{Jk} = \varphi_0^2/L_{k}$ for $k\in\{c, w, cw\}$ then use the sum trigonometric identity to expand the junction cosine terms
\begin{align}
-E_{Jk}\cos{(\hat\phi_k+\phi_{pk})} = -\frac{\varphi_0^2}{L_{k}}[\cos{(\hat\phi_k)}\cos{(\phi_{pk})}\\-\sin{(\hat\phi_k)}\sin{(\phi_{pk})}]
\end{align}
where $\hat\phi_{cw}=\hat\phi_w-\hat\phi_c$. Then we keep and relabel terms quadratic in $\hat\phi_k$:
\begin{equation}
\frac{\hat\phi_k^2\varphi_0^2}{L_{k}(\phi_{pk})} = \frac{\cos{(\phi_{pk})}\hat\phi_k^2\varphi_0^2}{L_{k}}.
\end{equation}
The quadratic term of $E_{d}\cos{(\hat\phi_d)}$ is simply $\hat\phi_d^2\varphi_0^2/L_{d}$. Using this procedure we may collect the coefficients of the identified terms to construct an effective inverse inductance matrix from Eq.~\ref{eq:charge}
\begin{align}
\check{L}^{-1} = \begin{pmatrix}
    \frac{1}{L_{d}} & 0 & 0\\
    0 & \frac{1}{L_{c}(\phi_{pc})}+\frac{1}{L_{cw}(\phi_{pcw})} & -\frac{1}{L_{cw}(\phi_{pcw})}\\
    0 & -\frac{1}{L_{cw}(\phi_{pcw})} & \frac{1}{L_{w}(\phi_{pw})}+\frac{1}{L_{w}(\phi_{pcw})}.
\end{pmatrix}
\end{align}

A rough estimate of the coupling--typically accurate to within $30\%$--can be obtained by substituting the harmonic oscillator transformation into the inductive coupling branch of the DTC and taking $\phi_{pcw} \rightarrow \phi_p$ and $\phi_{pc},\,\phi_{pw}\rightarrow 0$. The closure branch junction term may be expanded using trigonometric identities
\begin{equation}
\begin{split}
&\cos{(\hat\phi_w-\hat\phi_c+\phi_p)}=\\
&[\cos{(\hat\phi_w)}\cos{(\hat\phi_c)}
+\sin{(\hat\phi_w)}\sin{(\hat\phi_c)}]\cos{(\phi_p)}\\
&-[\sin{(\hat\phi_w)}\cos{(\hat\phi_c)}-\cos{(\hat\phi_w)}\sin{(\hat\phi_c)}]\sin{(\phi_p)},
\end{split}
\end{equation}
then we may collect terms proportional to $\hat\phi_w\hat\phi_c$
\begin{align}
-E_{Jcw}\hat\phi_w\hat\phi_c\cos{(\phi_p)},\label{eq:approxcoupling}
\end{align}
and finally apply the harmonic oscillator basis transformation
\begin{align}
\frac{\hbar\sqrt{Z_wZ_c}}{2L_{Jcw}}\cos{(\phi_p)}(a_w-a^{\dagger}_w)(a_c-a^{\dagger}_c)\label{eq:couplingHO}\\
\approx \hbar g_L(\phi_p)(a_w-a^{\dagger}_w)(a_c-a^{\dagger}_c).
\end{align}
Note that the $\sin{(\phi_w)}\cos{(\phi_c)}-\cos{(\phi_w)}\sin{(\phi_c)}$ terms thrown away in Eq.~\ref{eq:approxcoupling} would make a lesser, but still substantial, correction to the entangling interaction in Eq.~\ref{eq:couplingHO}.

Experimentally, we sinusoidally modulate $\phi_p(t) = \bar\phi_p + \tilde{\phi}_{p}(t)\sin{(\omega t)}$, where $\tilde{\phi}_p(t)$ is a dimensionless amplitude and $\bar\phi_p$ is the time averaged dimensionless flux. In the limit where $\tilde{\phi}(t)\ll 2\pi$ 
\begin{align}
\cos{(\phi_p)}=&\cos{\left(\bar\phi_p+\tilde{\phi}_p(t)\sin{(\omega_p t)}\right)}\nonumber\\ 
= &\cos{\left(\bar\phi_p\right)}\cos{(\tilde{\phi}_p(t)\sin{(\omega_p t)})}\nonumber\\
&-\sin{\left(\bar\phi_p\right)}\sin{(\tilde{\phi}_p(t)\sin{(\omega_p t)})}\nonumber\\
\approx &\cos{\left(\bar\phi_p\right)}-\tilde{\phi}_p(t)\sin{\left(\bar\phi_p\right)}\sin{(\omega_p t)}
\end{align}
From this expansion, we can estimate the time-independent and time-dependent parts of $g_L(t) = \bar g_{L} + \tilde{g}_{L}(t) \sin{(\omega_pt)}$
\begin{align}
\bar g_L &\approx \frac{\hbar\sqrt{Z_wZ_c}}{2L_{cw}}\cos{(\bar\phi_p)}\\
\tilde{g}(t) &\approx -\frac{\hbar\sqrt{Z_wZ_c}}{2L_{cw}}\tilde{\phi}_p(t)\sin{(\bar\phi_p)}
\end{align}

\section{Effective TED Hamiltonian derivation}
\label{sec:Hamiltonian_derivation}

Modeling the behavior of a TED can be computationally intensive since $Q_d$, $Q_c$, and $Q_w$ should have very different frequencies and there is at least one high-frequency driving term. In this section we will derive the effective stationary Hamiltonian used in the text, starting with a harmonic oscillator approximation of the exact Hamiltonian in Eq.~\ref{eq:charge}.
\begin{align}
\frac{H}{\hbar} ={}&
\sum_{q=\{d,c,w\}}[\omega_{q} q^{\dagger} q +\nu_q q^{\dagger 2} q^2]
\nonumber\\
& + g_{C} (d+ d^{\dagger}) (c+c^{\dagger}) \nonumber\\
&- g_{L}(t) (c- c^{\dagger}) (w-w^{\dagger})
\nonumber\\
& -i\Omega(t) \sin{(\omega_{\alpha} t)} (w-w^{\dagger}).
\label{eq:TEDHsupp}
\end{align}
To make the parametric drive stationary, we impose a rotating frame transformation $d \rightarrow de^{-i\omega_p t}$ and $c \rightarrow ce^{-i\omega_p t}$, resulting in the Hamiltonian
\begin{align}
\frac{H}{\hbar} ={}&
\Big[(\omega_{d}+\omega_p) d^{\dagger} d + (\omega_{c}+\omega_p) c^{\dagger} c + \omega_{w} w^{\dagger} w
\nonumber\\
& +\sum_{q=\{d,c,w\}}\nu_q q^{\dagger 2} q^2\nonumber\\
& + g_{C} (cde^{2i\omega_pt}+cd^{\dagger}+c^{\dagger}d+c^{\dagger}d^{\dagger}e^{-2i\omega_pt})
\nonumber\\
& - (\bar{g}_L+\tilde{g}_L(t)\sin{(\omega_p t)}) (ce^{i\omega_pt}- c^{\dagger}e^{-i\omega_pt}) (w-w^{\dagger})
\nonumber\\
& -i\Omega(t) \sin{(\omega_{\alpha} t)} (w-w^{\dagger})\Big].
\label{eq:TEDHrotframe}
\end{align}
We then apply the rotating wave approximation to eliminate fast rotating terms proportional to $e^{2i\omega_pt}$, resulting in the new Hamiltonian
\begin{align}
\frac{H}{\hbar} ={}&
\Big[(\omega_{d}+\omega_p) d^{\dagger} d + (\omega_{c}+\omega_p) c^{\dagger} c + \omega_{w} w^{\dagger} w
\nonumber\\
& +\sum_{q=\{d,c,w\}}\nu_q q^{\dagger 2} q^2\nonumber\\
& + g_{C} (cd^{\dagger}+c^{\dagger}d) + \tilde{g}_L(t)(c^{\dagger}w-cw^{\dagger})/2i
\nonumber\\
& -i\Omega(t) \sin{(\omega_{\alpha} t)} (w-w^{\dagger})\Big].
\label{eq:TEDHRWA}
\end{align}
Similarly, if there is a coherent drive we may perform the simultaneous transformations $d\rightarrow de^{-i\omega_{\alpha}t}$, $c\rightarrow ce^{-i\omega_{\alpha}t}$, and $w\rightarrow we^{-i\omega_{\alpha}t}$ to obtain
\begin{align}
\frac{H}{\hbar} ={}&
\Big[(\omega_{d}+\omega_p-\omega_{\alpha}) d^{\dagger} d + (\omega_{c}+\omega_p-\omega_{\alpha}) c^{\dagger} c 
\nonumber\\
&+ (\omega_{w}-\omega_{\alpha}) w^{\dagger} w + \sum_{q=\{d,c,w\}}\nu_q q^{\dagger 2} q^2\nonumber\\
& + g_{C} (cd^{\dagger}+c^{\dagger}d)
\nonumber\\
& + \tilde{g}_L(t)(c^{\dagger}w-cw^{\dagger})/2i + \Omega(t)(w+w^{\dagger})/2\Big].
\label{eq:TEDHRWAcoh}
\end{align}
In the next section we will eliminate the $Q_c$ mode to recover the equation in the main text.

\subsection{Adiabatic elimination of $Q_c$}
\label{sec:adiabatic_elimination}

The coupling qubit $Q_c$ mediates a virtual interaction between $Q_d$ and $Q_w$. This effective interaction may be computed using the Schrieffer-Wolff transformation 
\begin{align}
H' &= e^S H e^{-S}
\\
S &= \frac{g_C}{\omega_d-\omega_c}(d^{\dagger}c-c^{\dagger}d)+\frac{-\tilde{g}_L(t)/2i}{\omega_w-(\omega_c+\omega_p)}(w^{\dagger}c+c^{\dagger}w)
\end{align}
The Schrieffer-Wolff procedure is to divide a Hamiltonian into on-site and coupling terms: $H \approx H_0 + V$. Starting from Eq.~\eqref{eq:TEDHRWA} while neglecting the anharmonicity and coherent drive terms
\begin{align}
H_0 &= (\omega_{d}+\omega_p) d^{\dagger} d + (\omega_{c}+\omega_p) c^{\dagger} c + \omega_{w} w^{\dagger} w
\\
V &= g_{C} (cd^{\dagger}+c^{\dagger}d) + \tilde{g}_L(t)(c^{\dagger}w-cw^{\dagger})/2i
\\
H' &= H_0+V+[S,H_0]+\frac{1}{2}[S,V]+\mathcal{O}(V^3).
\end{align}
$S$ was chosen so that $V+[S,H_0] = 0$, leaving $\frac{1}{2}[S,V]$ as the lowest-order interaction term. Plugging in for $S$ and $V$ we obtain
\begin{align}
\frac{1}{2}[S,V] = &\left[\frac{-g_C\tilde{g}_L(t)}{4i(\omega_d-\omega_c)}+\frac{-g_C\tilde{g}_L(t)}{4i(\omega_w-(\omega_c+\omega_p))}\right]
\nonumber\\
&\times(w^{\dagger}d-wd^{\dagger})
\nonumber\\
&+\frac{-g_C^2}{(\omega_d-\omega_c)}(d^{\dagger}d-c^{\dagger}c)
\nonumber\\
&+\frac{-\tilde{g}_L^2(t)}{4(\omega_w-(\omega_c+\omega_p))}(w^{\dagger}w-c^{\dagger}c)
\end{align}
where the term proportional to $w^{\dagger}d-wd^{\dagger}$ is an effective interaction term mediated by a driving field. The terms proportional to $d^{\dagger}d$, $c^{\dagger}c$, and $w^{\dagger}w$ are corrections to the frequencies $\omega_d$, $\omega_c$, and $\omega_w$, respectively. In the resonant case $\omega_p = \omega_w-\omega_d$, the effective coupling simplifies to 
\begin{align}
\label{eq:geff}
H_{\textrm{int}}& = -ig_p(t)(d^{\dagger}w-dw^{\dagger})\\
g_p(t)&=\frac{g_C\tilde{g}_L(t)}{2(\omega_d-\omega_c)}
\end{align}

Transforming into the rotating frame of an external coherent drive--as done between Eq.~\eqref{eq:TEDHRWA} and Eq.~\eqref{eq:TEDHRWAcoh}--then adding back in the anharmonicity terms, we recover the equation used in the main text:
\begin{align}
    H/\hbar = {}&
    \Big[(\delta-\delta_p) d^{\dagger}d 
    + \delta w^{\dagger}w\label{eq:RWAsquared}\\
    &+\nu_d d^{\dagger 2} d^2 + \nu_ww^{\dagger 2} w^2\nonumber\\
    &-ig_p(t) \left( d^{\dagger} w - d w^{\dagger} \right)+\Omega(t)(w+w^{\dagger})/2 \Big]\nonumber\\
    \delta_p = {}& \omega_w-\omega_d\pm\omega_p\label{eq:deltasign}\\
    \delta = {}& \omega_w-\omega_{\alpha}.
\end{align}
The sign choice in Eq.~\ref{eq:deltasign} should minimize the magnitude of $\delta_p$.

\section{Optimizing for $Q_d$ coherence and ease of operation}
\label{sec:acStark}

Shifting of the frequency of $Q_w$ as a function of drive amplitude $\tilde{g}_L(t)$ complicates waveguide photon shaping and capture. While these distortions can be measured and corrected with the arbitrary waveform generator (AWG) on the DTC's flux drive line, it is often better to passively mitigate the shift.

In the main text we identify two origins to the frequency shift with $\tilde{g}_L(t)$. The first is an ac Stark shift $\tilde{g}_L(t)^2/4(\omega_d-\omega_c)$ of $\omega_w$ and the second is rectification from curvature in the dispersion of $\omega_w(\phi_p)$. The maximum desired value for $\tilde{g}_L(t)$ is given by $g_p = g_C \tilde{g}_L/2 (\omega_d-\omega_c) = \gamma/2$. Distortion of the outgoing photon wavefunction increases with the \textit{ratio} of the ac Stark shift of $\omega_w$ over $g_p$ and so $\tilde{g}_L\leq g_C/2$. For high coherence $Q_d$ it is useful to keep the coupling between $Q_d$ and $Q_c$ in the dispersive limit: $g_C \le 0.1|\omega_d-\omega_c|$. 

Taking the upper limit of the two inequalities in the paragraph above: we substitute $g_C = 0.1|\omega_d-\omega_c|$ and $\tilde{g}_L = g_C/2 = 0.1|\omega_d-\omega_c|/2$ into $g_C \tilde{g}_L/2 (\omega_d-\omega_c) = \gamma/2$. We can write the optimization condition in terms of the detuning
\begin{align}
0.005|\omega_d-\omega_c|\gg\gamma.
\end{align}

The above approach also reduces rectification by reducing the required amplitude of $\tilde{g}_L$. We can further reduce rectification by increasing the slope of $\omega_w(\phi_p)$ so we may modulate $\phi_p$ with a smaller amplitude to obtain the same $\tilde{g}_L(t)$. While increasing the value of $E_{Jcw}$ accomplishes this, it also increases many noise terms. We cap the value $E_{Jcw}\le0.2\,\textrm{min}(E_{Jc},E_{Jw})$.

\section{Calculating relaxation of the data qubit}
\label{sec:admittance}

Referring to the circuit in Fig.~\ref{fig:fig1}b, we use a perturbative approach for calculating an equivalent $1/RC$ decay rate for $Q_d$. This approach uses the admittance of the node associated with $Q_d$ to ground where the waveguide is replaced with a $50\,\Omega$ resistor to ground
\begin{align}
A &\parallel B = \frac{1}{\frac{1}{A}+\frac{1}{B}}\\
Y(\omega) &= i\omega C_d+\frac{1}{i\omega L_d} + i\omega C_{dc}\parallel \Bigg(i\omega C_c+\frac{1}{i\omega L_c}\nonumber\\
&+i\omega C_{dw} \parallel \left(i \omega C_w + \frac{1}{i\omega L_w}+i\omega C_v\parallel \frac{1}{50\,\Omega}\right)\Bigg)\\
\Gamma &\approx\frac{\Re(Y(\omega_d))}{C_d+C_{dc}}.
\end{align}
This approach is valid when the node of interest is only weakly hybridized with other nodes of the circuit. 

\section{Release and Detect Simulations}
\label{sec:SLH}

\begin{figure}
    \centering
    \includegraphics[]{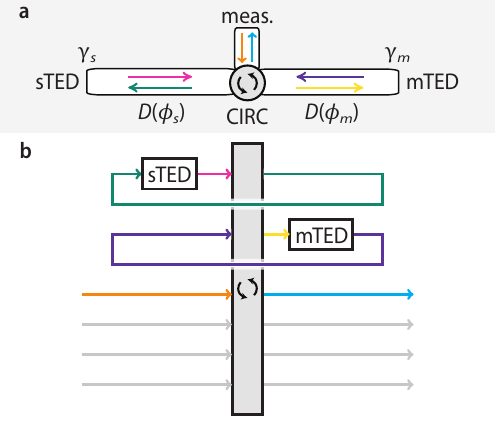}
    \caption{\textbf{(a)}~Schematic where a signal emitted by the sTED acquires a propagation phase $\phi_s$ before entering a circulator (CIRC), and subsequently acquires a phase $\phi_m$ while traveling to mTED for detection. \textbf{(b)}~The corresponding SLH diagram for this system. Three fictitious ports (indicated by grey arrows) are introduced
    to model circulator loss. The other signal path arrows are color-coded to emphasize the connection between \textbf{(a)} and \textbf{(b)}.}
    \label{fig:SLH_system}
\end{figure}

The release-and-detect experiment consists of three localized components (two TEDs and a circulator, Fig.~\ref{fig:SLH_system}a) networked together via itinerant bosonic modes. This network of devices is well-suited to a mathematical description via the SLH framework \cite{SLH1, Gough_2008, Combes_2017}.

In the SLH framework, localized components are described in terms of the triple $\slh{S}{L}{H}$ of scattering matrix, vector of Lindblad operators, and Hamiltonian. From this local description and the network topology, the SLH triple for the full network can be derived via three types of operation. The cascade operation between components $C_1$ and $C_2$, represented as $C_2 \triangleleft C_1$, sends the output from each port of component 1 into the corresponding input port of component 2. The concatenate operation between a $P_1$-port component and a $P_2$-port component, represented as $C_1 \boxplus C_2$, combines the two components into a single $(P_1 + P_2)$-port component. The feedback operation on a component $C$, represented as $[C]_{x\to y}$, feeds the output of port $x$ into input port $y$ of the same component. The master equation for a component with $P$ ports and SLH triple $\slh{S}{L}{H}$ is then given by
\begin{equation}
    \dot{\rho} = -\frac{\mi}{\hbar} [H, \rho] + \sum_{p=1}^P \mathcal{L}[L_p] \rho .
\end{equation}

A TED is a two-port component whose first port couples to right-moving modes and whose second port couples to left-moving modes. Its Hamiltonian $H_\mathrm{TED}$ is given by Equation~\eqref{eq:TEDHeff}, and its SLH triple is 
\begin{equation}
    \mathrm{TED} = \slh{\begin{bNiceMatrix} 1 &0 \\ 0 &1 \end{bNiceMatrix}}%
    {\begin{bNiceMatrix} \sqrt{\frac{\gamma}{2}} w \\ \sqrt{\frac{\gamma}{2}} w \end{bNiceMatrix}}%
    {H_\mathrm{TED}}.
\end{equation}
In our release-and-detect experiment, the source (sTED) and detector (mTED) both terminate the waveguide and thus are one-port components (as shown in Fig.~\ref{fig:SLH_system}b). The termination is modeled in the SLH framework via a feedback operation, and thus the $x$TED for $x \in \{\text{s, m}\}$ has SLH triple
\begin{equation}
    \label{eq:TEDfeedback}
    [\mathrm{TED}]_{1\to 2} = [\mathrm{TED}]_{2\to 1} = 
    \slh{1}%
    {\sqrt{2 \gamma_x} w_x}%
    {H_x}.
\end{equation}
The waveguide termination modifies the density of states in the waveguide, enhancing the coupling between the TED and the waveguide. This enhancement appears as a factor of 2 in Equation~\eqref{eq:TEDfeedback}, which we absorb into the definition of the decay rate $\gamma_x$ (this is now the experimental decay rate) to define the one-port $x$TED components as
\begin{equation}
    \mathrm{TED}_\text{one-port} = \slh{1}{\sqrt{\gamma_x} w_x}{H_x}.
\end{equation}

A lossy three-port circulator that directs inputs at port $p$ to output $p+1$ is represented by the SLH triple
\newcommand{\losscomp}{\bar{\eta}}
\begin{equation}
    \mathrm{CIRC} = \slh{
    \begin{bmatrix} 
    0 &0 &\losscomp &-\eta &0 &0 \\ 
    \losscomp &0 &0 &0 &-\eta &0 \\ 
    0 &\losscomp &0 &0 &0 &-\eta \\
    0 &0 &\eta &\losscomp &0 &0  \\
    \eta &0 &0 &0 &\losscomp &0  \\
    0 &\eta &0 &0 &0 &\losscomp  
    \end{bmatrix}}%
    {\begin{bmatrix} 0 \\ 0 \\ 0 \\ 0 \\ 0 \\ 0 \end{bmatrix}}%
    {0},
    \label{eq:CIRC}
\end{equation}
where $\eta \in [0, 1]$ is a loss parameter, $\losscomp = \sqrt{1-\eta^2}$, and the last three ports represent loss channels (see Fig.~\ref{fig:SLH_system}b).

The components are networked together via approximately one meter of coaxial cable. The timescale for signal propagation between components, $\textrm{distance} / v \sim \mathcal{O}(\text{ns})$, is significantly shorter than the timescale of the system dynamics, $2\pi / \gamma \sim \mathcal{O}(\mu\text{s})$. In this regime, and under the assumption that the cable is dispersionless and linear, it is appropriate to model signal propagation through the cable as a phase delay component $D(\phi) = \slh{e^{\mi \phi}}{0}{0}$. We take the phase due to propagation between the sTED and circulator to be $\phi_s$ and the delay due to propagation between the circulator and mTED to be $\phi_m$.

The full network for the release-and-detect experiment is then described by the SLH expression (where $I_n$ is an $n$-port identity component)
\begin{equation}
\begin{aligned}
    \Bigl[ &\left( I_1 \boxplus \text{mTED} \boxplus I_4 \right)
    \triangleleft \left( D(\phi_s) \boxplus D(\phi_m) \boxplus I_4 \right)
    \\
    &\triangleleft \text{CIRC}
    \triangleleft \left( D(\phi_s) \boxplus D(\phi_m) \boxplus I_4 \right)
    \\
    &\triangleleft \left( \text{sTED} \boxplus I_5 \right)
    \Bigr]_{1\to 1, 2\to 2}.
\end{aligned}
\label{eq:SLHmodel}
\end{equation}
The overall model can be interpreted as a sequence of cascaded modules corresponding to the signal path shown in Fig.~\ref{fig:SLH_system}a. In this expression, the first port corresponds to the waveguide connecting the sTED to the circulator, the second port corresponds to the coaxial cable connecting the circulator to the mTED, the third port corresponds to the measurement port of the circulator, and the remaining three ports are loss ports. In this formalism, the series cascade operator ($\triangleleft$) connects modules sequentially---with the signal path propagating from right to left---while the concatenation operator ($\boxplus$) arranges components in parallel. Thus, reading Eq.~\eqref{eq:SLHmodel} from right to left, the sTED emits a signal into port 1, where it travels through the waveguide and picks up a phase of $\phi_s$ before entering the circulator. The circulator redirects the signal (minus some loss) to port 2, where it propagates through the coaxial cable and picks up a phase of $\phi_m$ before entering the mTED. Due to the feedback operation, the output of the mTED is directed back into port 2 (at the right side of the expression), where it propagates through the coaxial cable and picks up a phase of $\phi_m$ before entering the circulator. The circulator redirects the signal (minus some loss) to port 3, from which the signal leaves the system.

This results in the master equation
\begin{equation}
\begin{aligned}
    \dot{\rho} = {}&-\frac{\mi}{\hbar} [H,\rho] 
    + \gamma_s \mathcal{L}[w_s] \rho + \gamma_m \mathcal{L}[w_m] \rho
    \\
    &+ \sqrt{(1-\eta^2) \gamma_s \gamma_m} \Bigl(\mathcal{L}_2\bigl[ w_s, e^{-\mi (\phi_s + \phi_m)} w_m \bigr] \rho 
    \\
    &+ \mathcal{L}_2\bigl[ e^{-\mi (\phi_s + \phi_m)} w_m, w_s \bigr] \rho \Bigr),
\end{aligned}
\end{equation}
where we define $\mathcal{L}_2[A, B] \rho = A \rho B^\dagger - \frac{1}{2} \{ A^\dagger B, \rho\}$ and the Hamiltonian is given by
\begin{equation}
\begin{aligned}
    H = {}&H_s + H_m 
    + \frac{1}{2} \sqrt{(1-\eta^2) \gamma_s \gamma_m} 
    \\
    &\times \Bigl( i e^{-i(\phi_s + \phi_m)} w_s^\dagger w_m 
    - i e^{i(\phi_s + \phi_m)} w_s w_m^\dagger \Bigr).
\end{aligned}
\end{equation}

For any output solely generated by the source qubit, the input-output relation~\cite{Eichler2011, Eichler2012, Lalumiere2013} has the following form
\begin{align}
\langle a_{\textrm{out}}\rangle = e^{i(\phi_s+\phi_m)}\sqrt{\frac{\gamma_s}{2}}\langle w_s\rangle + \sqrt{\frac{\gamma_m}{2}}\langle w_m\rangle\\
\langle w_s\rangle = \textrm{Tr}[\rho w_s]\\
\langle w_m\rangle = \textrm{Tr}[\rho w_m].
\end{align}

\section{Averaged measurements of the waveguide}
\label{sec:avgmeas}

\begin{figure}
    \includegraphics[width=3.3in]{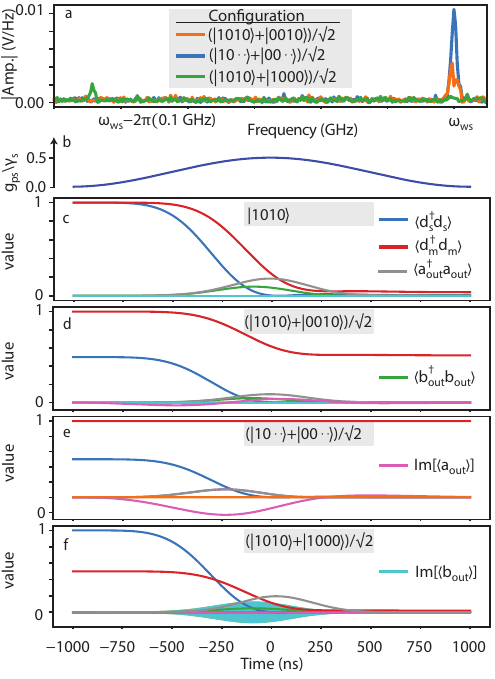}
    \caption{\textbf{(a)} The signal output along the waveguide is demodulated, measured by a digitizer, and digitally Fast Fourier transformed (FFT) for three initial states $|n_{ds}n_{ws}n_{dm}n_{dw}\rangle$ of the pitch-detect experiment. In \textbf{(b-f)} we show theory for the state evolution of the pitch-detect demonstration while neglecting loss in the waveguide. $|\langle a_\text{out}\rangle|$ and $|\langle b_\text{out}\rangle|$ are proportional to the experimental data in panel (a) observed at $\omega_{ws}/2\pi$ and $\omega_{ws}/2\pi-0.108(1)$~GHz, respectively. The simulation is in a frame that rotates at rate $\omega_{ws}$. \textbf{(b)} Shows the cosine envelope of $g_{ps}$, which releases the state stored in $Q_{ds}$. \textbf{(c)} Fock state emission and detection matching the pitch-detect demonstration in the main text. \textbf{(d)} Same as panel (c) where $Q_{ds}$ is instead prepared on the equator of its Bloch sphere. \textbf{e} Same as panel (d) where we tune $Q_{wm}$ away from $Q_{ws}$. \textbf{(f)} Same as panel (c) where we instead prepare $Q_{dm}$ on the equator of its Bloch sphere.
    }
    \label{fig:avgmeas}
\end{figure}

We put the waveguide emission seen in the measurement chain in context with theoretical quantum state populations of the joint sTED and mTED system. The state populations are structured as described in the main text $|n_{ds}n_{ws}n_{dm}n_{wm}\rangle$, where $n$ denotes the excitation level of each indexed transmon.

The amplitude and phase of a coherent state in a waveguide is given by the expectation value of the lowering operator $\langle a_\text{out}\rangle$. This value is proportional to the information retained after averaging a signal many times on a digitizer. We define our waveguide state using Appendix~\ref{sec:SLH} as $a_\text{out} = e^{i(\phi_s+\phi_m)}\sqrt{\gamma_s/2}\, w_s  +  \sqrt{\gamma_m/2}\, w_{m}^{(01)}$ and $b_\text{out} =  \sqrt{\gamma_m/2}\, w_{m}^{(12)}$ where $w_{m}^{(01)}$ is the part of the lowering operator $w_m$ that acts on the $0\leftrightarrow 1$ transition while $w_{m}^{(12)}$ acts on the $1\leftrightarrow 2$ transition. 
Figure~\ref{fig:avgmeas}a shows the fast Fourier transformation (FFT) of digitizer traces averaging over ten thousand measurements. It shows peaks proportional to $\langle a_\text{out}\rangle$ and $\langle b_\text{out}\rangle$ for different configurations of the pitch-detect demonstration.

We compare these experimental results to the model in Appendix~\ref{sec:SLH} using modulation of the envelope of $g_{ps}$ shown in Fig.~\ref{fig:avgmeas}b at frequency $\omega_{ps} = \omega_{ws}-\omega_{ds}$. The mTED is in a detection configuration such that the drive amplitude is $|g_{pm}| = \gamma_m/2$ and its frequency is $\omega_{pm} = \omega_{wm}-\omega_{dm}+\nu_{wm}$. The expectation values $\langle d_s^{\dagger}d_s \rangle$ and $\langle d_m^{\dagger}d_m \rangle$ track the populations of the data qubits $Q_{ds}$ and $Q_{dm}$, respectively. 
$\langle a_\text{out}^{\dagger}a_\text{out} \rangle$ and $\langle b_\text{out}^{\dagger}b_\text{out} \rangle$ are proportional to the itinerant microwave photon wavefunction power at the $0\leftrightarrow 1$ and $1\leftrightarrow 2$ transition frequencies, respectively. The real part of $\langle a_\text{out} \rangle$ is 0 for all times. We do not show $\Re[\langle b_\text{out} \rangle]$ because its carrier is simply $\pi/2$ out of phase with $\Im[\langle b_\text{out} \rangle]$.

The configuration $\ket{1010}$, corresponding to the main text, shows a Fock state emitted from the sTED that then de-excites the mTED. Note that $Q_{dm}$ does not go all the way to zero. Emission from the $1\leftrightarrow 2$ transition slightly leads emission from the $0\leftrightarrow 1$ transition. Preparing $Q_{ds}$ in a superposition means $\langle a_\text{out} \rangle$ can be non-zero as shown in Fig.~\ref{fig:avgmeas}d-e (pink curve). The amplitude of $\langle a_\text{out} \rangle$ is much smaller in Fig.~\ref{fig:avgmeas}d than  Fig.~\ref{fig:avgmeas}e because the mTED is measuring the waveguide state, projecting the waveguide state into a vacuum or Fock state. In Figure~\ref{fig:avgmeas}e we detune $\omega_{wm}$ from $\omega_{ws}$ so that the sTED's waveguide state is no longer measured. If instead we prepare $Q_{dm}$ in a superposition state, $\langle b_\text{out} \rangle$ is non-zero (Fig.~\ref{fig:avgmeas}f, cyan curve). Figure~\ref{fig:avgmeas}f is in the rotating frame of $\omega_{ws}$ causing $\langle b_\text{out} \rangle$ to have a carrier frequency $\nu_{wm}/2\pi$. Note that $\nu_{wm}$ differs from the value listed in Table~\ref{tab:params} because the experiment was reproduced in another cryostat (so we could add the waveguide measurement chain) and an avoided level crossing of $84.5(1)$~MHz, caused by a two level system (TLS), appeared $32(1)$~MHz above $\omega_{wm}/2\pi=5.7177(2)$~GHz. The TLS was not present in subsequent cryostat cooldowns.

We further checked the Fock state emission by setting $Q_{ds}$ and $Q_{dm}$ to either $\ket{0}$ or $\ket{1}$. We integrated over a single measurement of $a_{out}$ and $b_{out}$ after amplification, squared each, then averaged ten million of these measurements, the result of which is related~\cite{Eichler2011} to $\langle a_{out}^{\dagger}a_{out}\rangle$ and $\langle b_{out}^{\dagger}b_{out}\rangle$. These averages are then compiled in Table~\ref{table:Fock} where the result for the initial state $\ket{0000}$ was subtracted from the other configurations. We observe photons only if $Q_{ds}$ is initially excited. The last row of the table corresponds to the simulation in Fig.~\ref{fig:avgmeas}c where both $Q_{ds}$ and $Q_{dm}$ are initially excited, a photon is detected, and after amplification signatures of photons at both $\omega_{ws}$ and $\omega_{ws}-\nu_{wm}$ are observed. 

\begin{table}
\begin{tabular}{llll}
$Q_{ds}$ & $Q_{dm}$ & $0\leftrightarrow 1$ (arb.) & $1\leftrightarrow 2$ (arb.)\\
\hline
0 & 0 & reference & reference\\
1 & 0 & 1.1(2) & 0.1(2)\\
0 & 1 & -0.2(2) & -0.1(2)\\
1 & 1 & 2.1(2) & 1.5(2)
\end{tabular}
\caption{For each $Q_{ds}$ and $Q_{dm}$, ten million measurements were integrated, squared, then averaged. The average for $\ket{0000}$ was subtracted from the other configurations.}\label{table:Fock}
\end{table}

\section{Coherent scattering off $Q_w$}
\label{sec:scatQw}

Master equations are a convenient tool for modeling near-resonance coherent scattering with a qubit~\cite{Lalumiere2013}. We use this tool to perform the simulation in Fig.~\ref{fig:fig3}g. Including thermal terms, the master equation is
\begin{align}
\dot\rho = -\frac{i}{\hbar}[H, \rho] + \gamma(1+n_{\textrm{th}})\mathcal{L}[w] + \gamma n_{\textrm{th}}\mathcal{L}[w^{\dagger}]
\end{align}
where $n_{\textrm{th}}$ is the Bose-Einstein mode occupation at a given temperature and frequency.

We define here an RWA Hamiltonian that only includes the waveguide qubit and an external coherent drive term. Using the Hamiltonian Eq.~\eqref{eq:RWAsquared} instead can reveal how the scattering changes as a function of parametric drive power and frequency.
\begin{align}
H/\hbar = \delta w^{\dagger}w+\nu_ww^{\dagger 2}w^2 +i\Omega(w-w^{\dagger})/2\\
\Omega = 2\sqrt{\frac{2P_{\alpha}\gamma\omega_{\alpha}}{\omega_w}}\\
P_{\alpha} = \gamma \bar n
\end{align}
where $P_{\alpha}$ is the power of the coherent drive and $\bar n$ is the average number of coherent photons at frequency $\omega_{\alpha}$ and integrated over time $1/\gamma$. The output state of the waveguide can be written in terms of the input state with the following relation
\begin{align}
a_{\textrm{out}} = a_{\textrm{in}} + \sqrt{\frac{\gamma}{2}}w
\end{align}
where $a_{\textrm{in}}$ and $a_{\textrm{out}}$ are the lowering operators (with units of $\sqrt{\textrm{Hz}}$) for waveguide modes  that travel towards the qubit and away from the qubit, respectively. This input-output relation can be used to simulate various moments of the waveguide~\cite{Eichler2011, Eichler2012, Kannan2020, Kannan2023, Almanakly2025} including coherent and Fock state statistics. For a simulation of coherent scattering the simulation of interest is
\begin{align}
\langle a_{\textrm{out}}\rangle = \sqrt{\gamma \bar n} + \sqrt{\frac{\gamma}{2}}\langle w\rangle\\
\langle w\rangle = \textrm{Tr}[\rho w].
\end{align}
The equivalent experimental measurement can be performed with a two-port vector network analyzer.

\newpage

\section{Cryostat measurement setup}
\label{sec:measurementsetup}

\begin{figure*}[h!] 
    \includegraphics[width=\linewidth]{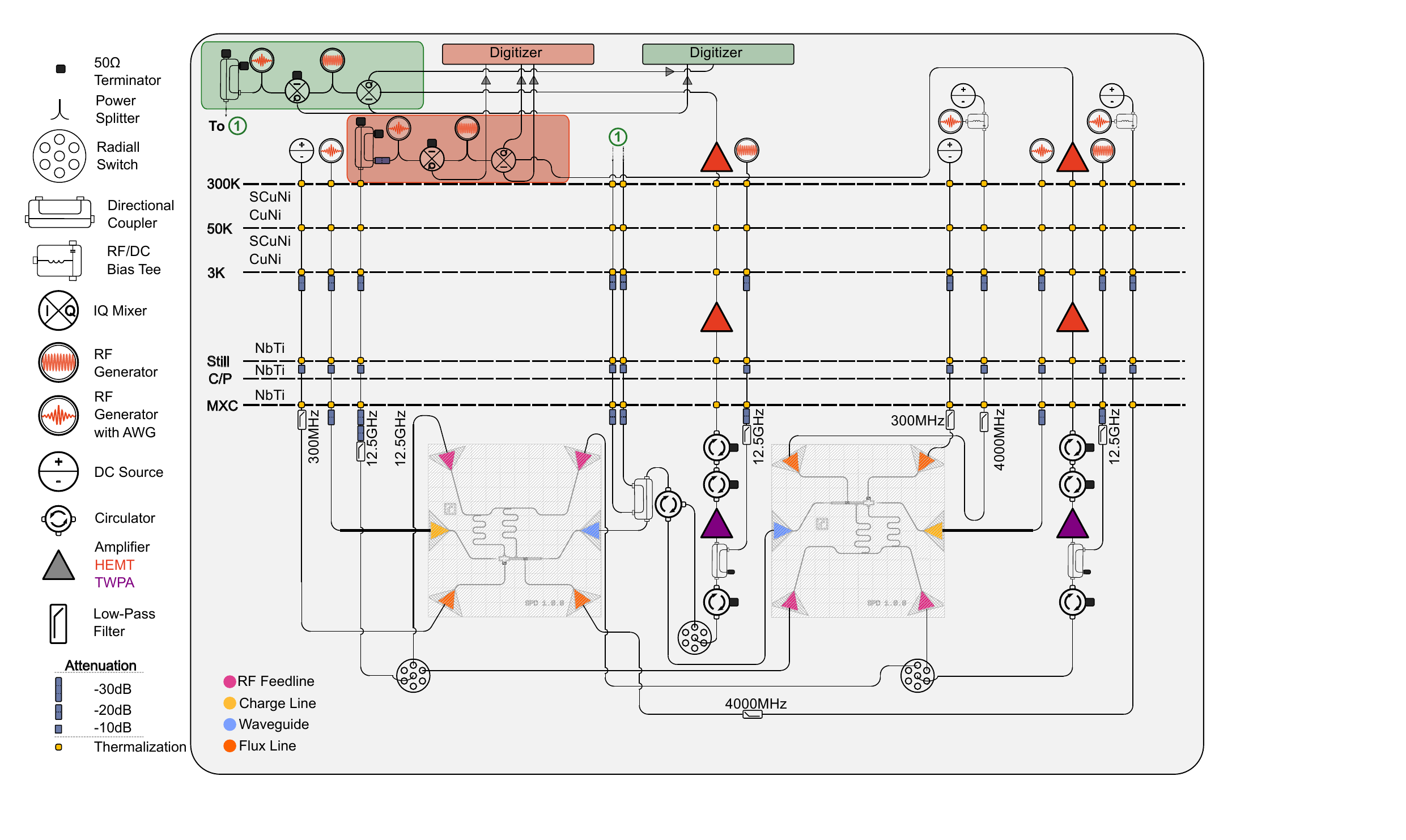}
    \caption{Diagrammatic representation of the room temperature and cryogenic setup. Subsystems are color-coded by their function: the heterodyne readout circuit used to measure qubits via their respective resonators is highlighted in red, while the readout chain used for waveguide backscatter measurements (as shown in Fig.~\ref{fig:fig3}f) is highlighted in green.
    }
    \label{fig:fridgelayout}
\end{figure*}

The cryostat measurement setup is shown in  Fig.~\ref{fig:fridgelayout} and components are listed in Table~\ref{tab:components}. Pulses were generated using Keysight M3202A arbitrary waveform generators (AWGs) inside of a Keysight M9019A chassis (not pictured), then routed to Rohde \& Schwarz (R\&S) SGS100A microwave frequency generators. Voltage sources (Yokogawa GS200) were either used in isolation on $Q_d$ flux lines, or coupled via room temperature bias tees (Mini-Circuits ZFBT-6GW+) to R\&S RF generators to parametrically drive while still allowing for modification of applied voltage bias. $Q_d$ flux lines were filtered by low pass filters (Mini-Circuits VLFX-300+) to prevent Purcell decay. Readout microwave drives were split into a drive/readout side using a power splitter (Marki Microwave PD0R618) to downconvert (Marki Microwave MMIQ0218LXPC) the received drive signal into the digitization band of the Keysight M3012A. Before digitization, the I/Q signal was amplified with a low-noise amplifier (SRS SR445A). Within the fridge, Radiall switches were used to switch between the different sample feed lines. The cryostat contained two sets of switches. Each drive side switches' center pin was used to send signals into the designated connection through 70~dB total attenuation in addition to a 12.5~GHz low-pass filter (RLC F-30-12.4-R). Measurement-side switches' center pins were attached to amplified lines consisting of a travelling-wave parametric amplifier (TWPA) at the mixing chamber (MXC), cryogenic high electron-mobility transistor (HEMT) at the 3K stage, and a room-temperature HEMT above the fridge. Dual-junction circulators were used between the sample and the TWPA to isolate from TWPA back-action, and between the TWPA and cryo-HEMT with similar reasoning. A directional coupler (Marki Microwave C10-0116) was used to couple the TWPA drive pump with the signal line. The circulator (LNF-CICIC4-12A) between the two samples was connected to a measurement-side Radiall switch distinct from the measurement line used for $Q_d$ readout. This allowed readout to be conducted in parallel with waveguide backscatter measurements. 

\begin{table*}[] 
 \renewcommand{\arraystretch}{1.5} 
 \newcolumntype{L}[1]{>{\raggedright\arraybackslash}p{#1}}
    \centering 
    \caption{Components in the Measurement Setup}
    \label{tab:components}
    \begin{NiceTabular}{|L{0.31\textwidth} |L{0.31\textwidth}| L{0.31\textwidth} |}
       \hline
         \textbf{Component} & \textbf{Manufacturer \& Model} & \textbf{Function} \\
       \hline
       \hline
          PXIe Chassis         & Keysight M9019A            & Hosts AWG and Digitizer modules \\
        Arbitrary Waveform Generator (AWG)  & Keysight M3202A            & Pulse sequence generation \\
RF generator &  Rohde \& Schwarz SGS100A               & Upconversion of AWG pulses to RF frequencies \\
DC Source            & Yokogawa GS200             & Static flux biasing for transmons \\
 Bias Tee             & Mini-Circuits ZFBT-6GW+    & Combines RF drive and DC flux bias on a single line \\
  Power Splitter       & Marki Microwave PD0R618              & Splits readout tone for I/Q downmixing reference \\
 I/Q Mixer        & Marki Microwave MMIQ0218LXPC          & Downconverts readout signal to baseband \\
Room Temperature  Preamplifier    & SRS SR445A 350 MHz    & Room temperature pre-amplification before digitization \\
 Digitizer            & Keysight M3012A  PXIe with FPGA           & I/Q signal acquisition and digitization \\
 RF Switch            & Radiall       R591762600             & Switches signal paths between different samples inside the fridge \\
12.5 GHz Low-Pass Filter     & RLC F-30-12.4-R   & Filters low-frequency signal lines  \\
 300 MHz Low-Pass Filter     & Mini-Circuits VLFX-300+    & Thermal noise filtering for DC flux lines  \\
10dB Broadband Directional Coupler  & Marki Microwave C10-0116             & Injects TWPA pump tone into the signal line \\
Circulator/Isolator  & Low Noise Factory LNF-CICIC4-12A             & Isolates sample from amplifier back-action and routes signal \\
 Traveling-Wave Parametric Amplifier (TWPA)                 &   MIT Lincoln Laboratory \& IARPA LogiQ program               & MXC stage, quantum-limited signal amplification \\
 Room temperature High Electron-Mobility Transistor (HEMT) &     LNF-LNR4\_8C\_SV            &  Readout line room-temperature amplifier \\
 Cryogenic HEMT  & Low Noise Factory LNF-LNC4\_8  & 4K stage cryogenic amplifier\\
        \hline
    \end{NiceTabular}
    \\
\end{table*}
\bibliographystyle{apsrev4-2}
\bibliography{Main,newbib}
\end{document}